\pdfoutput=1

\documentclass[11pt]{article}

\usepackage[]{acl}

\usepackage{times}
\usepackage{latexsym}

\usepackage[T1]{fontenc}

\usepackage[utf8]{inputenc}

\usepackage{microtype}

\usepackage{amsmath}
\usepackage{amsfonts}
\usepackage{amsthm}
\usepackage{booktabs}
\usepackage{multirow}
\usepackage{makecell}
\usepackage{subfigure}
\usepackage{enumitem}
\usepackage{graphicx}

\title{Denoising Neural Network for News Recommendation with Positive and Negative Implicit Feedback}

\author{Yunfan Hu\thanks{\ \ This work was done when Yunfan Hu was an intern at Tencent.} \\
  Santa Clara University \\
  \texttt{freddy@pku.edu.cn } \\\And
  Zhaopeng Qiu \\
  Tencent \\
  \texttt{zhaopengqiu@tencent.com} \\\AND
  Xian Wu\thanks{\ \ Xian Wu is the Corresponding Author.} \\
  Tencent \\
  \texttt{kevinxwu@tencent.com} \\}

\begin{document}
\maketitle
\begin{abstract}
News recommendation is different from movie or e-commercial recommendation as people usually do not grade the news. Therefore, user feedback for news is always implicit (click behavior, reading time, etc). Inevitably, there are noises in implicit feedback. On one hand, the user may exit immediately after clicking the news as he dislikes the news content, leaving the noise in his positive implicit feedback; on the other hand, the user may be recommended multiple interesting news at the same time and only click one of them, producing the noise in his negative implicit feedback. Opposite implicit feedback could construct more integrated user preferences and help each other to minimize the noise influence. Previous works on news recommendation only used positive implicit feedback and suffered from the noise impact. In this paper, we propose a {\bf D}enoising neural network for news {\bf R}ecommendation with {\bf P}ositive and {\bf N}egative implicit feedback, named DRPN. DRPN utilizes both feedback for recommendation with a module to denoise both positive and negative implicit feedback to further enhance the performance. Experiments on the real-world large-scale dataset demonstrate the state-of-the-art performance of DRPN.
\end{abstract}

\section{Introduction}

Online news platforms, such as Google News and Microsoft News, have attracted a large population of users~\cite{wu2020mind}. 
However, massive news articles emerging every day on these platforms make it difficult for users to find appealing content quickly~\cite{NPA}. To alleviate the information overload problem, recommender systems have become integral parts of these platforms.

A core problem in news recommendation is how to learn better representations of users and news~\cite{GNUD}. Early works include collaborate filtering (CF) based methods~\cite{das2007google}, content-based methods~\cite{ijntema2010ontology} and hybrid methods~\cite{de2012chatter} that combine the two. These methods usually have the cold start problem when being exposed to the sparsity
of user-item interactions~\cite{zhu2019dan}. Recently, deep learning methods have been proposed to learn better user and news representations. The techniques evolve from using recursive neural network~\cite{okura2017embedding}, attention mechanism~\cite{zhu2019dan, NRMS}, to graph neural network~\cite{wang2018dkn, GNUD, hu2020graph,10.1145/3511708}. These methods usually recommend news for users based on their historical feedback.

\begin{figure}[t]
  \centering
  \includegraphics[width=\linewidth]{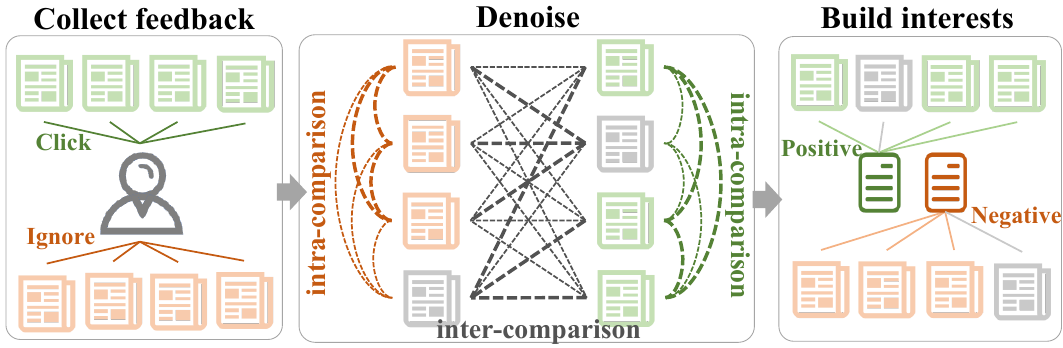}
  \caption{The idea to denoise the implicit feedback. Noises are found by conducting inter-comparison and intra-comparison and then reduced.} 
  \label{fig:de}
\end{figure}

Implicit feedback is more commonly collected than explicit feedback for news because the users usually do not grade the news. 
Hence, current news recommendation methods naturally use positive implicit feedback like click behavior as the historical feedback to model user interests.
However, there are gaps between positive implicit feedback and user real preferences~\cite{wang2018modeling}. 
For example, the click behaviors do not fully reflect the user's preferences. 
The user may exit the news immediately after clicking, which will involve a noise in the positive feedback.
Additionally, some news that users did not click, may also attract them later.
Ignoring them also impacts the recommendation performance. 
Our observation is that using both positive and negative implicit feedback can better model user interests.
Besides, positive and negative implicit feedback can help to denoise each other by conducting inter-comparison and intra-comparison.
If a news story in one feedback sequence is more similar to the news in the opposite feedback sequence rather than the news in the same sequence, it is very likely that this news story constitutes noise.
We can remove this news when building user interests. 
This idea is shown in Figure~\ref{fig:de}.

In this paper, we propose the {\bf D}enoising neural network for news {\bf R}ecommendation with {\bf P}ositive and {\bf N}egative implicit feedback, named DRPN. 
It first introduces a news encoder to represent the news in two implicit feedback sequences. 
Then two parallel aggregators are used to extract user representations from both positive and negative historical feedback: (1) \textit{content-based aggregator}, which selects the informative news in the feedback sequences to represent the user;
(2) \textit{denoising aggregator}, which finds and reduces the noises in the feedback sequences.
In addition to the semantic information, we introduce a graph neural network to incorporate the collaborative information to further enrich the user representation.
Finally, the user and candidate news representations are used to predict the clicking probability. 
The contributions of this paper are summarized as follows:
\begin{itemize}
    \item We propose a novel neural news recommendation approach DRPN which jointly models both positive and negative implicit feedback sequences to represent the user to improve recommendation performance.
    \item In DRPN, to minimize the impacts of the noises in the implicit feedback, the denoising aggregators are designed to refine the two feedback sequences and can help to further improve the recommendation performance.
    \item The experiments on the large-scale real-world dataset demonstrate that DRPN achieves state-of-the-art performance.
\end{itemize}

\section{Related Works}

\subsection{Recommendation with Multi-type Feedback}
Few works notice the noise problem in the implicit feedback. \cite{zhao2018recommendations, liu2020hypernews} use multiple types of feedback to improve recommendation. However, they ignore the noise in the implicit feedback. 
\cite{wang2018modeling} notices the noise problem but it fails to use the meaningful semantic information in the news. 
\cite{CRPS, xie2020deep, recsysdenoise} use the explicit feedback (such as reading time and like/dislike behaviors) to help denoise the implicit feedback. 
However, the explicit feedback is harder to collect than the implicit feedback.
Differently, DRPN only depends on the implicit feedback (click and non-click behaviors) to conduct the denoise to better model the user preferences.

\subsection{Graph Neural Network}
Recently, graph neural networks (GNN) have received wide attention in many fields~\cite{wu2020comprehensive}. The convolutional GNN can learn powerful node representations by aggregating the neighbors' features. Recently, some works have attempted to leverage the graph information to enhance the representations learning for news recommendation with GNNs. \cite{wang2018dkn} uses entities in news to build a knowledge graph and use the entity embeddings to improve the model performance. \cite{GERL} combines the one- and two-hop neighbor news and users to enrich the representations of the candidate news and user, respectively.
However, these methods also depend on the positive implicit feedback to model user representations and ignore the noise problem.

\section{Problem Formulation}
The news recommendation problem in our paper can be illustrated as follows.
Let \(\mathcal{U}\) and  \(\mathcal{R}\) denote the entire user set and news set.
The feedback matrix for the users over the news is denoted as $\mathcal{Z}\in \mathbb{R}^{l_u\times l_r}$, where $z_{u,r}=1$ means user $u$ gives a positive implicit feedback to news $r$ (e.g., \(u\) clicks \(r\)), $z_{u,r}=-1$ means user $u$ gives a negative implicit feedback to news $r$ (e.g., \(u\) sees \(r\) but ignores it), and $z_{u,r}=0$ means no feedback.
$l_u$ and $l_r$ denote the numbers of the users and news, respectively.
For each specific user, his historical positive feedback sequence $[p_1, ..., p_{l_p}]$ and negative feedback sequence $[n_1, ..., n_{l_n}]$ can be gathered from the feedback matrix $\mathcal{Z}$, where $p_i, n_j\in \mathcal{R}$.

Given the feedback matrix  \(\mathcal{Z}\), the goal is to train a model \(\mathcal { M }\) (i.e., GRPN).
For each new pair of user and candidate news \((u\in \mathcal{U}, r\in \mathcal{R})\), we can use \(\mathcal { M }\) to estimate the probability that \(u\) would like to click \(r\).

\section{Framework}

\begin{figure}[t]
  \centering
  \includegraphics[width=\linewidth]{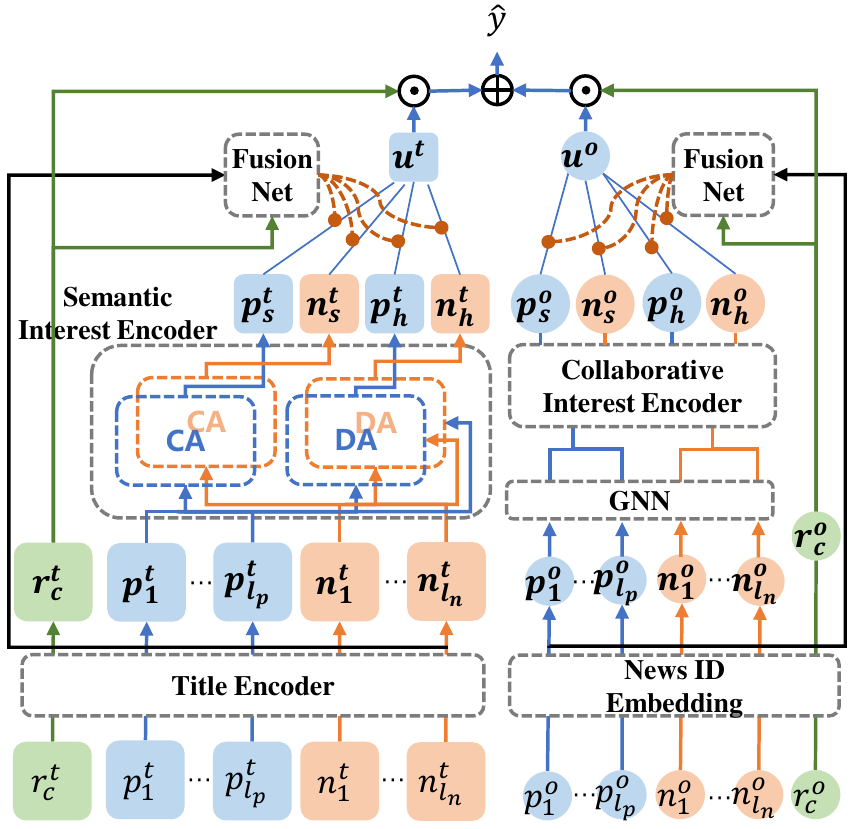}
  \caption{The DRPN framework.} 
  \label{fig:framework}
\end{figure}

Figure \ref{fig:framework} shows the architecture of DRPN. 
It first employs the title encoder and id embedding layer to represent all news in two feedback sequences and the candidate news.
Then two separate encoders are employed to extract the user semantic interest and collaborative interest information from both positive and negative implicit feedback sequences. 
Next, two fusion nets combine multiple interest representations to represent the user. 
Finally,  we use the user and candidate news representations to estimate the clicking probability.
We will detail each component in the following subsections.


\subsection{Input}
The inputs of the DRPN contain six parts: the titles of positive feedback sequence $[p^t_1,..., p^t_{l_p}]$, the titles of negative feedback sequence $[n^t_1, ..., n^t_{l_n}]$, the candidate news title $r_c^t$, the IDs of positive feedback sequence $[p^o_1,..., p^o_{l_p}]$, the IDs of negative feedback sequence $[n^o_1,..., n^o_{l_n}]$, and candidate news ID $r_c^o$.

For each news title $t$, we convert its every word \(w\) to a \(d\)-dimensional vector \(\mathbf{w}\) via an embedding matrix $\mathbf{E}_{W}\in \mathbb{R}^{l_w\times d}$, where $l_w$ is the vocabulary size and \(d\) is the dimension of word embedding. Then, the title $t$ is transformed into a matrix \(\mathbf{T}\).

For each news ID $o$, we also convert it to a \(d\)-dimensional vector $\mathbf{o}$ via an embedding matrix $\mathbf{E}_{O}\in \mathbb{R}^{l_r\times d}$. In this manner, we can encode all news to obtain $\mathbf{P}^o = [\mathbf{p}^o_1,..., \mathbf{p}^o_{l_p}]$ ($\mathbf{p}^o_i\in \mathbb{R}^{d}$),  $\mathbf{N}^o = [\mathbf{n}^o_1,..., \mathbf{n}^o_{l_n}]$ ($\mathbf{n}^o_i\in \mathbb{R}^{d}$), and $\mathbf{r}_c^o\in \mathbb{R}^{d}$.

\begin{figure}[t]
  \centering
  \includegraphics[width=\linewidth]{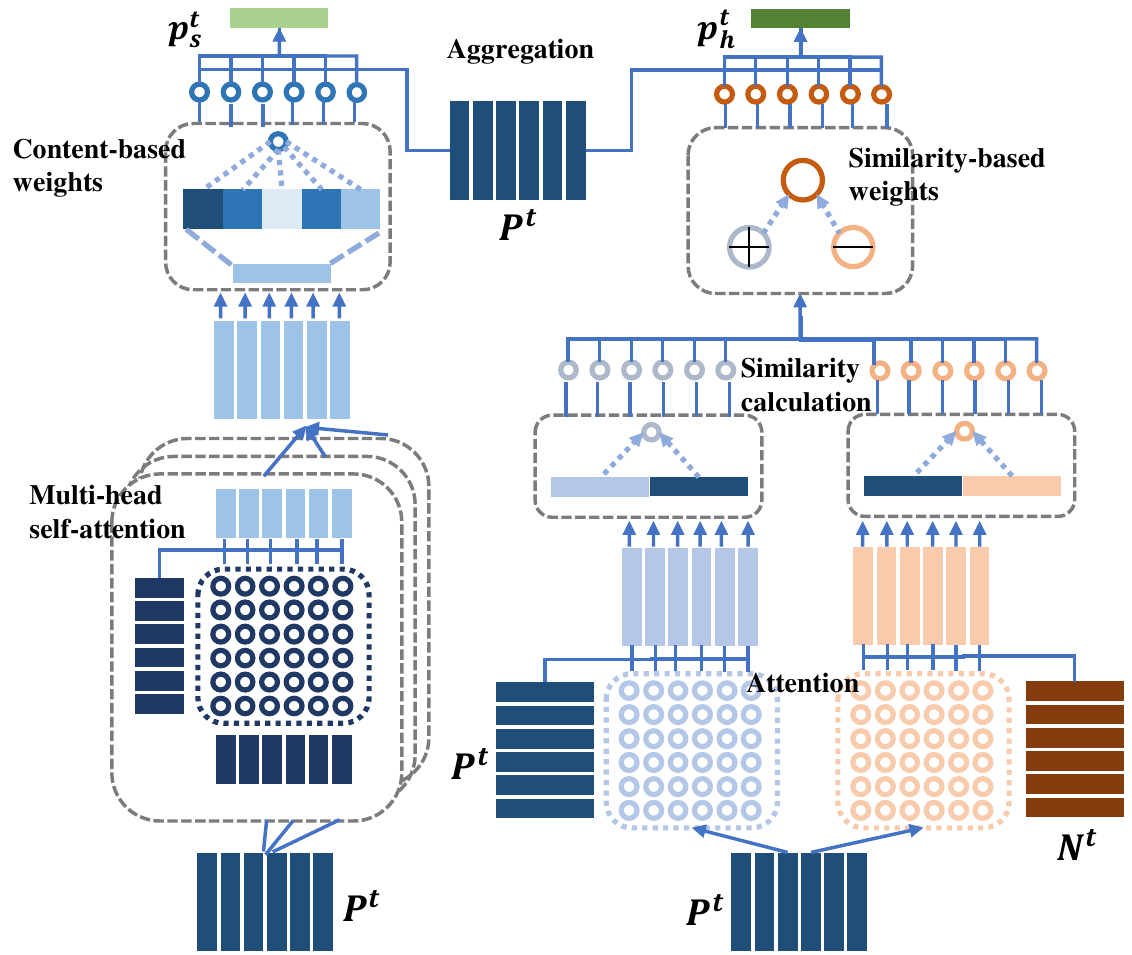}
  \caption{The content-based aggregator (CA) on the left and the denoising aggregator (DA) on the right in semantic interest encoder. They encode positive preferences for the user. \emph{(best viewed in color)}} 
  \label{fig:enco}
\end{figure}

\subsection{Title Encoder}\label{subsection:te}
The title encoder can extract the sentence-level semantic representation of the news title. 
It contains two sub-layers.
We take the title \(\mathbf{T}\) as an example to detail the encoding process. 

The first sub-layer is a multi-head self-attention layer, which can model the contextual representation of each word.
Given three input matrices $\mathbf{Q}\in \mathbb{R}^{l_q\times d}$, $\mathbf{K}\in \mathbb{R}^{l_v\times d}$ and $\mathbf{V}\in  \mathbb{R}^{l_v\times d}$, the attention function is defined as:
\begin{equation}
\label{eq:att}
\texttt{Attn}(\mathbf{Q}, \mathbf{K}, \mathbf{V}) = \text{softmax}(\mathbf{Q}\mathbf{K}^\top/{\sqrt{d}}) \mathbf{V}
\end{equation}
Multi-head self-attention layer $\texttt{MH}(\cdot, \cdot, \cdot)$ will further project the input to multiple semantic subspaces and capture the interaction information from multiple views:
\begin{equation}
\label{eq:mh}
\begin{split}
    & \texttt{MH}(\mathbf{Q}, \mathbf{K}, \mathbf{V}) = [head_1;...;head_{l_h}]\mathbf{W}^I \\
    & head_i = \texttt{Attn}(\mathbf{Q} \mathbf{W}^Q_i, \mathbf{K} \mathbf{W}^K_i, \mathbf{V} \mathbf{W}^V_i)
\end{split}
\end{equation}
where $\mathbf{W}^Q_i$, $\mathbf{W}^K_i$, $\mathbf{W}^V_i\in \mathbb{R}^{d\times d/l_h}$ and $\mathbf{W}^I\in \mathbb{R}^{d\times d}$ are the parameters to learn. $l_h$ is the number of heads. $[;]$ denotes the column-wise concatenation for matrices.

Moreover, we employ the residual connection and layer normalization function $\texttt{LN}$ defined in~\cite{Ba2016LayerN} to fuse the original and contextual representations: $\widetilde{\mathbf{T}} = \texttt{LN}(\mathbf{T} + \texttt{MH}(\mathbf{T}, \mathbf{T}, \mathbf{T}))$.

The second sub-layer is a gated aggregation layer~\cite{DBLP:conf/coling/QiuWF20}. It will select the important words to generate an informative title representation. The gated mechanism is employed to decide the weight of each words. Given the word embedding matrix $\widetilde{\mathbf{T}}$, its sentence-level semantic representation $\mathbf{t}$ is calculated as follows:
\begin{equation}
\label{eq:aggregate}
\begin{split}
    \mathbf{t} &= \texttt{Aggregate}(\widetilde{\mathbf{T}}) \\ 
     &= \left( \text{softmax}(\text{tanh}(\widetilde{\mathbf{T}}\mathbf{W}_a + \mathbf{b}_a)\mathbf{W}_g) \right)^{\top} \widetilde{\mathbf{T}}
\end{split}
\end{equation}
where $\mathbf{W}_{a}\in\mathbb{R}^{d\times d'}$, $\mathbf{b}_a\in\mathbb{R}^{d'}$ and $\mathbf{W}_{g}\in\mathbb{R}^{d'\times 1}$ are trainable parameters.

Finally, we can use the title encoder to model the titles of all news in two user feedback sequences to obtain $\mathbf{P}^t = [\mathbf{p}^t_1,..., \mathbf{p}^t_{l_p}]$ and $\mathbf{N}^t = [\mathbf{n}^t_1,..., \mathbf{n}^t_{l_n}]$.
For the candidate news, we can also obtain its title representation \(\mathbf{r}_c^t\) via the same title encoder.

\subsection{Semantic Interest Encoder}
The titles of the news which the user interacted usually reflect the user's interests.
Hence, we can learn user interest representations by encoding the semantic representations of the news.
As is shown in Figure~\ref{fig:enco}, the semantic interest encoder leverages two aggregators, \textit{content-based aggregator} (CA) and \textit{denoising aggregator} (DA), to extract user preferences from both positive and negative feedback sequences.

\subsubsection{Content-based Aggregator}
Different news have different informativeness when representing users.
For example, sport news are more informative than weather news in modeling user personality, since the latter are usually browsed by most users.
The content-based aggregator (CA) will first evaluate the importance of different news in the feedback sequence from the content view and then aggregate the important news to represent the user.
It contains two sub-layers.

The first one is a multi-head self-attention layer, which can enhance the news representations by capturing their interactions.
For the positive feedback sequence $\mathbf{P}^t$, the multi-head self-attention layer generates $\widetilde{\mathbf{P}}^t = \texttt{LN}(\mathbf{P}^t + \texttt{MH}(\mathbf{P}^t, \mathbf{P}^t, \mathbf{P}^t))$. The $\texttt{MH}$ is define in Eq.(\ref{eq:mh}) with independent parameters and the $\texttt{LN}$ is the layer normalization function. 

The second sub-layer is a gated aggregation layer that has the same structure as the one defined in Eq.(\ref{eq:aggregate}).
For $\widetilde{\mathbf{P}}^t$, it can select the more informative news to generate the user representation: $\mathbf{p}^t_{s} = \texttt{Aggregate}(\widetilde{\mathbf{P}}^t)$. 
We also use the content-based aggregator to generate another user representation from the negative feedback sequence $\mathbf{N}^t$, $\mathbf{n}^t_{s}$.

\subsubsection{Denoising Aggregator}
Denoising aggregator will conduct what we call a refining operation, which aims to mitigate the impacts of the noises in the feedback when modeling the user interests.
Intuitively, if news clicked by the user is more semantically relevant with the news in the positive feedback sequence, this news is more likely the user true preference.
Otherwise, if it is more semantically relevant with the news in the negative feedback sequence, it is more likely a noise for representing the user interest.
As shown in Figure~\ref{fig:enco}, for each news in the positive feedback sequence, we will conduct the intra-comparisons with the news in the positive sequence and inter-comparisons with the news in the negative sequence to decide its weight when representing the user.
This module contains three sub-layers.

The first sub-layer is an intra-attention layer. For news $\mathbf{p}_j^t\in \mathbf{P}^t$, this layer uses it as the query to aggregate all news in $\mathbf{P}^t$ except $\mathbf{p}_j^t$ by the attention mechanism to obtain the sequence-level representation, $\hat{\mathbf{p}}^t_j = (\sum_{i\neq j} \alpha^h_{ji}\mathbf{p}^t_i)\mathbf{W}^h_{v}$, where
\begin{equation}
\label{eq:msa}
    \alpha^h_{ji} = \frac{\exp\left((\mathbf{p}^t_j\mathbf{W}^h_{q})(\mathbf{p}^t_i\mathbf{W}^h_{k})^\top/\sqrt{d}\right)}{\sum\limits_{e\neq j}\exp\left((\mathbf{p}^t_j\mathbf{W}^h_{q})(\mathbf{p}^t_e\mathbf{W}^h_{k})^\top/\sqrt{d}\right)}
\end{equation}
$\mathbf{W}^h_q$, $\mathbf{W}^h_k$, $\mathbf{W}^h_v\in \mathbb{R}^{d\times d}$ are learnable parameters. 

The second sub-layer is an inter-attention layer.
For $\mathbf{p}_j^t$, this layer uses it as the query to aggregate its relevant news in the negative feedback sequence $\mathbf{N}^t$ by the attention mechanism.
\begin{equation}
\label{eq:cross}
    \hat{\mathbf{n}}_j^t=\texttt{Attn}(\mathbf{p}_j^t\mathbf{W}^h_{q}, \mathbf{N}^t\mathbf{W}^{h^{'}}_{k}, \mathbf{N}^t\mathbf{W}^{h^{'}}_{v})
\end{equation}
where $\mathbf{W}^{h^{'}}_k$, $\mathbf{W}^{h^{'}}_v\in\mathbb{R}^{d\times d}$ are learnable parameters. 

The third sub-layer is a gated aggregation layer.
The weight of the news $\mathbf{p}_j^t$ is decided by the semantic similarities between $\mathbf{p}_j^t$ and two sequence-level representations, $\hat{\mathbf{p}}^t_j$ and $\hat{\mathbf{n}}^t_j$.
\begin{equation}
\label{eq:denoise}
\begin{split}
   & s^p_j = \text{tanh}([\mathbf{p}_j^t;\hat{\mathbf{p}}_j^t]\mathbf{W}^u_1 + \mathbf{b}^u_1)\mathbf{W}^u_2 + b^u_2\\
   & {s}^n_j = \text{tanh}([\mathbf{p}^t_j;\hat{\mathbf{n}}_j^t]\mathbf{W}^u_3 + \mathbf{b}^u_3)\mathbf{W}^u_4 + b^u_4\\
   & \alpha_j = \frac{\exp (s^p_j - \text{ReLU}(\gamma) * {s}^n_j)}{\sum_{i=0}^{i=l_p} \exp (s^p_i - \text{ReLU}(\gamma) * {s}^n_i)} \\
\end{split}
\end{equation}
where $\mathbf{W}^u_1$, $\mathbf{W}^u_3\in \mathbb{R}^{2d\times d}$, $\mathbf{W}^u_2$, $\mathbf{W}^u_4\in \mathbb{R}^{d\times 1}$, $\mathbf{b}^u_1$, $\mathbf{b}^u_3\in \mathbb{R}^{d}$, $b^u_2$, $b^u_4\in \mathbb{R}$ and $\gamma$ are learnable parameters. 
Then, this layer will aggregate all news according to their weights to obtain the denoised representatioin, $\mathbf{p}_h^t = \sum_{j=0}^{j=l_p} \alpha_j \mathbf{p}_j^t$.

For the negative feedback sequence $\mathbf{N}^t$, we take a dual denoising process to obtain its final representation $\mathbf{n}_h^t$.

\subsection{Graph Neural Network}
If two news, $r_i$ and $r_j$, are co-clicked by the user $u_1$ and $r_i$ is also clicked by $u_2$, $u_2$ may also prefers $r_j$ based on the idea of the collaborative filtering.
Hence, we can further enrich the user interest representations by modeling the collaborative information.
Like the knowledge graph, we build a collaborative graph \(\mathcal{G}=\{(r_i, r_j)|r_i, r_j\in \mathcal{R}\}\) over the news set $\mathcal{R}$ based on the co-clicking relationships in the historical feedback matrix $\mathcal{Z}$.
$(r_i, r_j)$ indicates they are neighbors in the graph and have been clicked by the same user.
To incorporate the collaborative information, we employ the graph transformer neural network~\cite{shi2020masked} to model the news in the user feedback sequence.

First, for each news node $\mathbf{r}^o$ in $\mathbf{P}^o$ and $\mathbf{N}^o$, we compute the attention weights between it and its neighbors $\mathcal{N}(\mathbf{r}^o)$ in $\mathcal{G}$. $\mathcal{N}(\mathbf{r}^o)$ denotes the neighbor set of node $\mathbf{r}^o$. Take its neighbor $\mathbf{r}^o_k$ ($k\in \mathcal{N}(\mathbf{r}^o)$) as an example, the attention weight between $\mathbf{r}^o$ and $\mathbf{r}^o_k$ at the $m$-th head is calculated by
\begin{equation*}\label{eq:trans}
    \alpha^m_{k} = \frac{\exp\left((\mathbf{r}^o\mathbf{W}^g_{m,1})(\mathbf{r}^o_k\mathbf{W}^g_{m,2})^\top/\sqrt{\hat{d}}\right)}{ \sum\limits_{q\in \mathcal{N}(\mathbf{r}^o)}\exp\left((\mathbf{r}^o\mathbf{W}^g_{m,1})(\mathbf{r}^o_q\mathbf{W}^g_{m,2})^\top/\sqrt{\hat{d}}\right)}
\end{equation*}
where $\mathbf{W}^g_{m,*}\in \mathbb{R}^{d\times d/l_h'}$ are learnable parameters. $l_h'$ is the number of heads and $\hat{d}$ is equal to $d/l_h'$. 

Next, each news node will aggregate the information of its neighbors from multiple heads according to the attention weights. For the node $\mathbf{r}^o$, the representation aggregated from its neighbors is:
\begin{equation}
    \hat{\mathbf{r}}^o = \bigg\|_{m=1}^{m=l_h'} \left\{ \sum_{k\in \mathcal{N}(\mathbf{r}^o)}\alpha^m_{k}\mathbf{r}^o_k\mathbf{W}^g_{m,3} \right\}
\end{equation}
where $\{\mathbf{W}^g_{m,3}\in \mathbb{R}^{d\times d/l_h'}\}_{m=1}^{m=l_h'}$ are trainable parameters. $\|$ denotes the concatenation operation for $l_h'$ heads.

Finally, we will update the representation of each node by fusing its aggregated and original representations.
\begin{equation}
\label{eq:fuse}
\begin{split}
    \widetilde{\mathbf{r}}^o &= \texttt{Fuse}(\mathbf{r}^o, \hat{\mathbf{r}}^o) \\
    &= \sigma( [\mathbf{r}^o;\hat{\mathbf{r}}^o]\mathbf{W}^f_1)\odot \text{tanh}( [\mathbf{r}^o;\hat{\mathbf{r}}^o]\mathbf{W}^f_2)
\end{split}
\end{equation}
where $\mathbf{W}^f_1$, $\mathbf{W}^f_2\in \mathbb{R}^{2d\times d}$ are learnable parameters. $\odot$ denotes the element-wise multiplication operation. $\sigma$ is the sigmoid function. 

We can use this graph neural network to encode all news in the user positive and negative feedback sequences to obtain $\widetilde{\mathbf{P}}^o = [\widetilde{\mathbf{p}}^o_1,..., \widetilde{\mathbf{p}}^o_{l_p}]$ and $\widetilde{\mathbf{N}}^o = [\widetilde{\mathbf{n}}^o_1,..., \widetilde{\mathbf{n}}^o_{l_n}]$.

\subsection{Collaborative Interest Encoder}
The module aims to model user interests by aggregating the representations of two feedback sequences encoded by the graph neural network layer, which have incorporated the collaborative information.
The structure of the collaborative interest encoder is similar to that of the semantic interest encoder and also contains two aggregators, a content-based aggregator and a denoising aggregator.
The denoising aggregators have the same structure as the one in the semantic interest encoder.
The only structural difference between two content-based aggregators of two encoders is that there is no multi-head self-attention operation in the content-based aggregator of the collaborative interest encoder.
This is because the context information is already propagated by the graph neural work, which has a similar effect with the multi-head self-attention. 

The inputs of this encoder are the positive sequence representation $\widetilde{\mathbf{P}}^o$ and the negative sequence representation $\widetilde{\mathbf{N}}^o$.
The content-based aggregator will generate two user representations, $\mathbf{p}^o_{s}$ and $\mathbf{n}^o_{s}$, based on two sequence representations, respectively.
Similarly, the denoising aggregator will denoise two sequences and generate two user representations $\mathbf{p}^o_{h}$ and $\mathbf{n}^o_{h}$.

\subsection{Fusion Net}
There are two fusion nets as shown in Figure~\ref{fig:framework}. 
They are used to fuse multiple user interest representations extracted by two interest encoders to form a comprehensive user representation.
For different user-candidate news $(u, r)$ pairs, the fusion net dynamically allocates different weights for different interest representations.
Two fusion nets have similar structures but different parameters.
We take the one for the semantic interest encoder as an example to detail the fusion process.

The fusion net first represents the $(u, r)$ pair.
It should mitigate the effect of two interest encoders and independently calculate the weights for the output representations of two encoders.
Hence, it uses the outputs of the title encoder to represent $(u, r)$, $\mathbf{f}^t = [\mathbf{u}^t_f;\mathbf{r}_c^t]$, where $\mathbf{u}^t_f=\texttt{Aggregate}([\mathbf{P}^t|\mathbf{N}^t])$. $\mathbf{P}^t$, $\mathbf{N}^t$ and $\mathbf{r}_c^t$ are the title representations of the news in user positive and negative feedback sequences and the candidate news extracted by the title encoder. $[|]$ denotes the row-wise concatenation for matrices.

Then, this module leverages four different fully connected layers to calculate the weights for four representations extracted by the semantic interest encoder (i.e., $\mathbf{p}^t_{s}$, $\mathbf{n}^t_{s}$, $\mathbf{p}^t_{h}$ and $\mathbf{n}^t_{h}$).
For example, the weight of $\mathbf{p}^t_{s}$ is calculated by
\begin{equation}
\label{eq:fusion}
    \beta^p_{s} = \text{tanh}(\mathbf{f}^t\mathbf{W}^{p, s}_1 + \mathbf{b}^{p, s}_1)\mathbf{W}^{p, s}_2 + b^{p, s}_2
\end{equation}
where $\mathbf{W}^{p, s}_1\in \mathbb{R}^{2d\times d}$, $\mathbf{W}^{p, s}_2\in \mathbb{R}^{d\times 1}$, $\mathbf{b}^{p, s}_1\in \mathbb{R}^{d}$, $b^{p, s}_2\in \mathbb{R}$ are learnable parameters.
The weights, $\beta^n_{s}$, $\beta^p_{h}$ and $\beta^n_{h}$, of the representations $\mathbf{n}^t_{s}$, $\mathbf{p}^t_{h}$ and $\mathbf{n}^t_{h}$ can be calculated by the same way in Eq.(\ref{eq:fusion}).

Finally, the user content-view representation is calculated by
\begin{equation}
    \mathbf{u}^t = \beta^p_{s}*\mathbf{p}^t_{s} + \beta^n_{s}*\mathbf{n}^t_{s} + \beta^p_{h}*\mathbf{p}^t_{h} + \beta^n_{h}*\mathbf{n}^t_{h}
\end{equation}

Another fusion net is used to fuse four interest representations extracted by the collaborative interest encoder and has a similar structure with the above one.
The only difference is that it uses the outputs of the news ID embedding layer to represent the $(u,r)$ pair, $\mathbf{f}^o = [\mathbf{u}^o_f;\mathbf{r}_c^o]$, where $\mathbf{u}^o_f=\texttt{Aggregate}([\mathbf{P}^o|\mathbf{N}^o])$. 
Then, it uses $\mathbf{f}^o$ to calculate the respective weights $\theta^p_{s}$, $\theta^n_{s}$, $\theta^p_{h}$ and $\theta^n_{h}$ for four interest representations $\mathbf{p}^o_{s}$, $\mathbf{n}^o_{s}$, $\mathbf{p}^o_{h}$ and $\mathbf{n}^o_{h}$ by the same way in Eq.(\ref{eq:fusion}).
The final user graph-view representation is calculated by
\begin{equation}
    \mathbf{u}^o = \theta^p_{s}*\mathbf{p}^o_{s} + \theta^n_{s}*\mathbf{n}^o_{s} + \theta^p_{h}*\mathbf{p}^o_{h} + \theta^n_{h}*\mathbf{n}^o_{h}
\end{equation}

\subsection{Prediction}

Following~\cite{NRMS}, the clicking probability score \(\hat{y}\) is computed by the inner product of the user representation and the candidate news representation: $\hat{y} = {\mathbf{u}^t}^\top \mathbf{r}_c^t + {\mathbf{u}^o}^\top \mathbf{r}_c^o$,
where ${\mathbf{u}^t}^\top \mathbf{r}_c^t$ stands for the score calculated from title information and ${\mathbf{u}^o}^\top \mathbf{r}_c^o$ stands for the score calculated from collaborative information.

\subsection{Training}
Following~\cite{NRMS}, for each positive sample, we randomly select $l_k$ negative samples from the same user to construct a $l_k + 1$ classification task. 
Each output of the DRPN for a classification sample is like \([\hat{y}^+, \hat{y}^-_1, ..., \hat{y}^-_{l_k}]\), where \(\hat{y}^+\) denotes the clicking probability score of the positive sample and the rest denote the scores of the \(l_k\) negative samples.
We define the training loss (to be minimized) as follows.
\begin{equation}
    \mathcal{L} = -\sum_{i\in \mathcal{P}} \log ( \frac{\exp(\hat{y}_i^+)}{\exp(\hat{y}_i^+) + \sum\limits_{j=1}\limits^{j={l_k}} \exp(\hat{y}^-_{i,j})} )
\end{equation}
where $\mathcal{P}$ denotes the set of positive samples.

\section{Experiment}

\subsection{Dataset}

\begin{table}
\centering
\resizebox{1.0\columnwidth}{!}{
\begin{tabular}{ |l|r|l|r| } 
 \hline
 \textbf{\# users} & 654870 & \textbf{\# avg. titles words} & 12.66 \\
 \hline
 \textbf{\# news} & 104153 & \textbf{\# positive samples} & 1048414 \\ 
 \hline
 \textbf{\# words} & 54292 & \textbf{\# negative samples} & 25145229 \\ 
 \hline
 \multicolumn{3}{|l|}{\textbf{\# avg. positive feedback sequence length}} & {4.14}\\
 \hline
 \multicolumn{3}{|l|}{\textbf{\# avg. negative feedback sequence length}} & {96.80}\\
 \hline
\end{tabular}
}
\caption{Statistics of the dataset.}
\label{table:data}
\end{table}

There is no off-the-shelf dataset in which the user profile includes both positive and negative historical feedback sequences.
Therefore, we use MIND~\footnote{https://msnews.github.io/} dataset (its original user profile only contains positive feedback) to re-build one to conduct the experiments. 
The original MIND dataset contains the user impression logs. 
An impression log records the news displayed to a user when visiting the news website homepage at a specific time, and the click behaviors on the news list.
We re-build the dataset based on the MIND's impression logs as follows: (1) Select the impression logs of the first 5 days of the original training set. Then we add the news that a user has seen but did not click to his negative feedback sequence, and add the news he clicked to his positive feedback sequence. In this manner, the user profile includes both positive and negative historical feedback sequences; (2) Training set: the impression logs of 6-th day of the original training set; (3) Validation set: the first 10\% chronological impression logs of the original validation set; (4) Testing set: the last 90\% chronological impression logs of the original validation set.

The training, validation, and testing sets use the same user profile built in Step (1).
Since the user profiles are only built in Step (1) which is ahead of Step (2)-(4), there is no label leakage to validation and testing sets.
Moreover, same as the original MIND dataset, the re-built dataset also has 44.6\% users of validation set and 48.7\% users of test set that are not shown in the re-built training set.
Table~\ref{table:data} shows some statistics of the re-built dataset.

\subsection{Baseline Approaches and Metrics}
We evaluate the performance of DRPN by comparing it with several baseline methods, including: 
(1) \textit{LibFM}~\cite{rendle2012factorization}, factorization machine (FM); 
(2) \textit{DeepFM}~\cite{guo2017deepfm}, which combines the FM and neural networks; 
(3) \textit{DKN}~\cite{wang2018dkn}, which uses the CNN to fuse the entity and word embeddings to learn news representations;
(4) \textit{LSTUR}~\cite{LSTUR}, which uses the GRU to model short- and long-term interests from the click history;
(5) \textit{NPA}~\cite{NPA}, which introduces the attention mechanism to select important words and news;
(6) \textit{DEERS}~\cite{zhao2018recommendations}, which uses GRU to encode positive and negative feedback sequences;
(7) \textit{DFN}~\cite{xie2020deep}, a factorization-machine based network which uses transformers to encode both positive and negative feedback sequences to enhance performance;
(8) \textit{GERL}~\cite{GERL}, which constructs user-news graph to enhance the performance;
(9) \textit{NAML}~\cite{NAML}, which uses multi-view learning to aggregate different kinds of information to represent news;
(10) \textit{NRMS}~\cite{NRMS}, which uses multi-head self-attention to learn news and user representations;
(11) \textit{NAML + TCE}, which incorporates the denoising training strategy TCE~\cite{wang2021denoising} into NAML;
(12) \textit{NRMS + TCE}, which improves NRMS by using TCE.

Following the previous news recommendation work~\cite{wu2020mind,NRMS,GERL}, we use AUC, MRR, nDCG@5, and nDCG@10 scores as our evaluation metrics.

\subsection{Implementation Details}
For DRPN, the representation dimension $d$ is set to 300. We use the \textit{GloVe.840B.300d}~\cite{pennington2014glove} as the pre-trained word embeddings. 
The maximum title length is set to 15. 
The lengths of feedback sequences $l_p$ and $l_n$ are set to 30 and 60. 
Padding and truncation are used to keep sequence and word numbers the same. 
More implementation details are provided in Appendix~\ref{appendix:detail} and the computation complexity is provided in Appendix~\ref{appendix:cost}.  

For NRMS, DKN, LSTUR, NPA, and NAML, we use the official code and settings~\footnote{https://github.com/microsoft/recommenders}. For others, we reimplement them and set their parameters based on the experimental setting strategies reported by their papers. 

For fair comparisons, all methods only use the news ID, title, category and subcategory as features.
The validation set was used for tuning hyperparameters and the final performance comparison was conducted on the test set.

\begin{table}[ht]
\centering
\resizebox{1.0\columnwidth}{!}{
\begin{tabular}{ l|c|c|c|c } 
 \toprule
 \textbf{Models} & \textbf{AUC} & \textbf{MRR} & \textbf{nDCG@5} & \textbf{nDCG@10}\\ 
 \midrule
 LibFM & 60.48 & 26.38 & 27.75 & 34.63 \\
 DeepFM & 62.18 & 27.26 & 29.08 & 35.68 \\
 DKN & 64.00 & 28.98 & 31.49 & 38.22 \\
 LSTUR & 65.31 & 30.31 & 33.34 & 39.86 \\
 NPA & 64.35 & 29.61 & 32.88 & 39.23 \\
 DEERS & 65.29 & 30.78 & 33.78 & 40.09 \\
 DFN & 63.11 & 29.14 & 31.88 & 38.33 \\
 GERL & 64.08 & 29.34 & 32.50 & 38.96 \\
 NAML & 65.84 & 30.60 & 33.89 & 40.23 \\
 NRMS & 65.46 & 30.73 & 33.78 & 40.13 \\
 NAML + TCE & 65.95 & 30.66 & 33.93 & 40.52\\
 NRMS + TCE & 65.84 & 31.58 & 34.93 & 41.26\\
 \midrule
\textbf{DRPN} & \textbf{67.30} & \textbf{32.68} & \textbf{36.27} & \textbf{42.33} \\
 \bottomrule
\end{tabular}
}
\caption{Performance comparison of all methods. Best results are highlighted in bold.}
\label{tab:MIND}
\end{table}

\begin{figure*}[t]
  \centering
  \includegraphics[width=\linewidth]{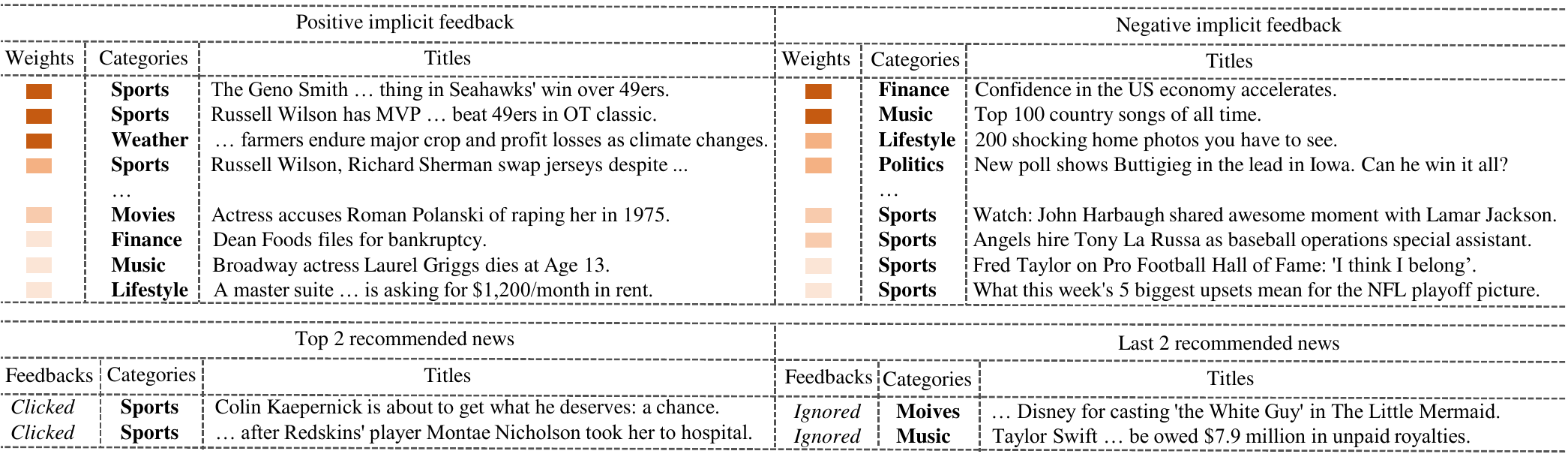}
  \caption{Visualization of the attention weights for an example user's feedback in the denoising aggregator and the recommendation results for him in validation dataset. A darker color indicates a larger attention weight. } 
  \label{fig:case}
\end{figure*}

\subsection{Performance Evaluation}
The experimental results of all models are summarized in Table~\ref{tab:MIND}.
We make the following observations from the results.
First, our proposed model, DRPN, outperforms all baselines on the news recommendation datasets.
Second, among all baselines, the methods which use the deep neural networks to model the news (i.e., DKN, NPA, LSTUR, DFN, DEERS, NAML, GERL, NRMS) perform better than the feature-based methods (e.g., LibFM and DeepFM). 
This performance improvement should be attributed to better news representation methods.
Among the deep neural methods, NRMS+TCE achieves the best performance by using two level multi-head self-attention to learn user representations and using TCE to denoise the negative samples.
Third, among two baselines that use both positive and negative feedback, DFN performs worse than DEERS.
The reason may be that original DFN depends on the explicit feedback but the experimental dataset only contains implicit feedback.
Compared with NAML, DEERS has a competitive performance even if its news encoder is a simple pooling layer.
This also proves the effectiveness of the negative implicit feedback.

\begin{table}
\centering
\resizebox{\columnwidth}{!}{
\begin{tabular}{ l|c|c|c|c } 
 \toprule
 \textbf{Models} & \textbf{AUC} & \textbf{MRR} & \textbf{nDCG@5} & \textbf{nDCG@10}\\ 
 \midrule
 \textbf{DRPN} & \textbf{67.30} & \textbf{32.68} & \textbf{36.27} & \textbf{42.33} \\
 \midrule
 DRPN-D & 66.51 & 31.09 & 35.40 & 41.63 \\
 DRPN-G & 66.82 & 31.90 & 35.34 & 41.62 \\
 DRPN-DG & 66.14 & 31.17 & 34.57 & 40.97 \\
 DRPN-N & 66.38 & 31.39 & 34.79 & 41.13 \\
 DRPN-P & 65.89 & 31.11 & 34.38 & 40.59 \\
 \bottomrule
\end{tabular}
}
\caption{Performance comparison of all variants of DRPN. Best results are highlighted in bold.}
\label{tab:ablation}
\end{table}

\subsection{Ablation Study}
To highlight the individual contribution of each module, we use the following variants of DRPN to run an ablation study:
(1) \textit{DRPN-D}, which removes the denoising aggregator;
(2) \textit{DRPN-G}, which removes the knowledge graph part;
(3) \textit{DRPN-DG}, which removes the knowledge graph part and the denoising aggregator;
(4) \textit{DRPN-N}, which only uses the positive feedback;
(5) \textit{DRPN-P}, which only uses the negative feedback.

The results are shown in Table \ref{tab:ablation}. First, DRPN-D and DRPN-G perform worse than DRPN, proving the effectiveness of the designed denoising module and the collaborative graph.
Second, the results of DRPN-N and DRPN-P indicate the effectiveness of negative and positive feedback, respectively. 
Third, even without deliberately designing, by using both positive and negative implicit feedback, DRPN-DG can achieve competitive performance compared with the strongest baseline NRMS+TCE.
This further proves the effectiveness of the negative feedback.

\subsection{Case Study}
To intuitively illustrate the effectiveness of the denoising aggregator, we sample a user and visualize his historical feedback attention weights in the denoising aggregator of the semantic interest encoder.
The upper part of the Figure~\ref{fig:case} shows the attention weights and ranks the news in descending order of the attention weight.
We can find in positive feedback sequence, the top 4 news are about sports and weather and the last 4 news are about music, movie, finance, and lifestyle.
Meanwhile, in negative feedback sequence, the top 4 news are about finance, music, politics, and lifestyle, and the last 4 news are all about sports. 
This indicates that the denoising aggregator believes that the user likes sports, and dislikes the topics such as finance, music, movies, politics, and lifestyle.
As shown in the lower part of Figure \ref{fig:case}, based on the predicted user preferences, we can see DRPN prefers to recommend the sports news for this user.
Moreover, in the validation data, we can observe that this user clicks the top 2 recommended news and ignores the last 2 news.
It suggests the user preference extracted by the denoising aggregator is consistent with the user's real behaviors.
In summary, the visualization results indicate the denoising module can better capture the user's real preferences by conducting the inter- and intra- comparisons between the positive and negative implicit feedback sequences.

\section{Conclusion}
In this paper, we propose a novel deep neural news recommendation model DRPN. In DRPN, we design two aggregators to extract user interests from both positive and negative implicit feedback.
The content-based aggregator focuses on the contents in the news representations and the denoising aggregator aims to mitigate the noise impact commonly existing in the implicit feedback.
Besides, apart from the title information, DRPN also exploits the collaborative information by the graph neural network to further improve the recommendation performance.
Experimental results on a large-scale public dataset demonstrate the state-of-the-art performance of DRPN.
The further study results also show the effectiveness of the denoising module.

\bibliography{custom}
\bibliographystyle{acl_natbib}

\appendix
\section{Implementation Details}
\label{appendix:detail}
The head number $l_h$ in multi-head self-attention is set to 6. 
The hidden size $d'$ in the gated aggregation layer is set to 200. 
The head number in graph neural network $l_h'$ is set to 2. 
The negative sampling ratio $l_k$ is set to 4.
When preparing data for graph neural network, we only input sub-graph that contains nodes in the user feedback sequences. 
Moreover, we pick the maximum 5 neighbor nodes for each node $r$ in user feedback sequences, which are most frequently co-clicked with $r$.

\section{Computation Complexity}
\label{appendix:cost}
The time complexity of the title encoder is $\mathcal{O}(L^2d+Ld^2)$, where $L$ is the title length and $d$ is the embedding size.
The time complexity of each interest encoder is $\mathcal{O}((l_p+l_n)d^2 + (l_p^2+l_n^2+(l_p+l_n)^2)d)$ where $l_p$ and $l_n$ are the lengths of positive and negative feedback sequences.
The time complexity of GNN is $\mathcal{O}(|\mathcal{G}|d)$, where $|\mathcal{G}|$ denotes the number of edges that existed in collaborative graph.
Hence, The overall time cost is $\mathcal{O}((l_p+l_n)(Ld^2+L^2d)+ (l_p^2+l_n^2+(l_p+l_n)^2 + |\mathcal{G}|)d)$.

During the inference phase, we can compute the news representations in advance and the computation complexity will be $\mathcal{O}( (l_p^2+l_n^2+(l_p+l_n)^2)d)$.

\section{Discussion}
\subsection{Limitations}
In this paper, to better learn the representations, our method refines the historical behaviors of the user by the denoising manner.
There are still some potential directions to further improve our approach.
First, since the user profile in the experimental dataset only contains the historical behaviors and has no basic information (e.g., gender and age), our current approach doesn't support these features but they are widely used in practice.
After these features are ready, we can convert them to embeddings and fuse them with the semantic interest representations obtained by two interest encoders to better represent the user.
Second, the news generally contains many forms of features except for the title (such as the cover image and author information) and our approach will explore how to involve more features to better represent the news.

\subsection{Potential Risks}
Our approach is based on the collaborative filtering, which may lead to that all of recommended news are similar to what the user has seen.
This is a common problem faced by the majority of recommender systems.
The concentration of a large number of similar information may narrow users' perspective and result in an
imbalance on the personal information structure~\cite{DBLP:journals/access/LiW19h}.
Our method can combine with some rule/human-based strategies (such as popularity based recommendation) to improve the recommendation diversity to alleviate this problem.

\end{document}